\documentclass[useAMS,usenatbib]{mn2e}
\usepackage{times}
\usepackage{graphics,epsfig}
\usepackage{graphicx}
\usepackage{amssymb}
\usepackage{txfonts}

\newcommand{\kms}{km\,s$^{-1}$} 
\newcommand{\Ts}   {\ensuremath{{\rm T}_{\rm s}}}

\newcommand{\Tk}   {\ensuremath{{\rm T}_{\rm k}}}
\newcommand{\Tb}   {\ensuremath{{\rm T}_{\rm B}}}

\newcommand{\nhi}  {\ensuremath{N}_{\rm HI}}
\newcommand{\nhiot}{\ensuremath{N_{\rm HI, OT}}}
\newcommand{\nhits}{\ensuremath{N_{\rm HI, ISO}}}

\newcommand{\cm}{cm$^{-2}$}
\newcommand{\hi}{H{\sc i}}

\newcommand{\hii}{H{\sc i}\,21cm}

\title[Accurate measurement of the H{\sc i} column density]{Accurate measurement of the 
H{\sc i} column density from {H{\sc i}~21\,cm} absorption-emission spectroscopy}
\author[Chengalur et al.]{Jayaram N. Chengalur$^{1}$\thanks{E-mail: 
chengalu@ncra.tifr.res.in~(JNC)},
Nissim Kanekar$^{1}$ and Nirupam Roy $^{2}$\\
$^{1}$National Centre for Radio Astrophysics, TIFR, Post Bag 3, Ganeshkhind, Pune 411 007, India\\
$^{2}$Max-Planck-Institut f\"{u}r Radioastronomie, Auf dem H\"{u}gel 69, D-53121, Bonn, Germany}

\begin{document}
\date{Accepted yyyy month dd. Received yyyy month dd; in original form yyyy month dd}


\maketitle

\label{firstpage}

\begin{abstract}
We present a detailed study of an estimator of the \hi\ column density, based on a combination of \hii\ 
absorption and H{\sc i}\,21cm emission spectroscopy. This ``isothermal'' estimate is given by 
$\nhits = 1.823 \times 10^{18} \int \left[ \tau_{\rm tot} \times {\rm T_B} \right] /  
\left[ 1 - e^{-\tau_{\rm tot}} \right] {\rm dV}$, where $\tau_{\rm tot}$ is the total
H{\sc i}\,21cm optical depth along the sightline and ${\rm T_B}$ is the measured brightness temperature.
We have used a Monte Carlo simulation to quantify the accuracy of the isothermal estimate by 
comparing the derived $\nhits$ with the true H{\sc i} column density $N_{\rm HI}$.
The simulation was carried out for a wide range of sightlines, including gas in different temperature 
phases and random locations along the path. We find that the results are statistically insensitive 
to the assumed gas temperature distribution and the positions
of different phases along the line of sight. The median value of the ratio of the true H{\sc i} column 
density to the isothermal estimate, $N_{\rm HI}/{N_{\rm HI, ISO}}$, is within a factor of 2 of unity while 
the 68.2\% confidence intervals are within a factor of $\approx 3$ of unity, out to high H{\sc i} column 
densities, $\le 5 \times 10^{23}$\,cm$^{-2}$ per 1~km~s$^{-1}$ channel, and 
high total optical depths, $\le 1000$. The isothermal estimator thus provides a significantly better 
measure of the H{\sc i} column density than other methods, within a factor of a few of the true value even 
at the highest columns, and should allow us to directly probe the existence of high H{\sc i}
column density gas in the Milky Way.

\end{abstract}

\begin{keywords}
ISM: general -- radio lines: ISM
\end{keywords}

\section{Introduction}
\label{sec:intro}

The neutral atomic hydrogen column density $\nhi$ is an important input to our understanding
of the interstellar medium (ISM). For example, it determines whether the gas is 
predominantly ionized (for $\nhi \lesssim 10^{17}$~\cm, in the intergalactic medium), predominantly 
neutral (for $\nhi \gtrsim 2 \times 10^{20}$~\cm, in typical gas clouds in galaxies), or predominantly 
molecular (for $\nhi \gtrsim 10^{22}$~\cm, in compact molecular clouds). It serves as the reference 
for estimates of various interesting quantities such as gas metallicities and abundances, is 
required to derive the gas spin temperature, and is the basic input for models of gas clouds. Accurate 
estimates of $\nhi$ are thus critical for ISM studies.

There are two standard approaches towards measuring $\nhi$ in Galactic clouds. The first is based on 
absorption spectroscopy of stars and quasars in the Lyman-$\alpha$ line, which develops wide Lorentzian
wings for typical sightlines through the Milky Way (or external galaxies), and whose equivalent width 
is directly related to the \hi\ column density. Such damping wings are easily detectable with modern 
optical spectrographs for $\nhi \gtrsim 10^{19}$~\cm, and offer accurate $\nhi$ measurements for 
sightlines that do not contain much dust. Unfortunately, the presence of significant amounts of dust 
along a sightline causes obscuration of the background star/quasar and, further, the amount of dust 
obscuration correlates with the total hydrogen column density in the Milky Way. As a result, it is very 
difficult to use Lyman-$\alpha$ spectroscopy to measure $\nhi$ along high column density Galactic 
sightlines, with $\nhi \gg 10^{21}$~\cm. 

The second approach to $\nhi$ measurements is via \hii\ emission studies, which directly measure
the brightness temperature $\Tb$ of the emission. For optically-thin \hii\ emission, the \hi\ column 
density is proportional to $\Tb$, even when the emission arises from multiple gas ``clouds'' with 
different temperatures. The advantage of this method is that it is 
easy to detect \hii\ emission along any Galactic sightline with today's telescopes. Further, 
significant progress has recently been made in correcting for stray radiation, received through the 
telescope sidelobes \citep[e.g.][]{kalberla05,bajaja05}. At low to moderate column densities, 
$\nhi < 10^{21}$~\cm, comparisons between $\nhi$ estimates from the Lyman-$\alpha$ 
absorption and \hii\ emission approaches have typically yielded excellent agreement, to within 10\% 
\citep[e.g.][]{dickey90,wakker11}.

Unfortunately, the relation between $\Tb$ and $\nhi$ is not straightforward for the general case of 
arbitrary \hii\ optical depth. When the optical depth is significant, one has to know both the 
location of different emitting components along the sightline and their individual optical depths 
and spin temperatures to infer $\nhi$ from the measured $\Tb$. Assuming that the gas is optically thin 
only yields a lower limit on the \hi\ column density. 


Sightlines with high \hi\ column densities are also the ones that tend to have high \hii\ optical 
depths. It is thus difficult to accurately estimate $\nhi$ for such sightlines using either 
Lyman-$\alpha$ absorption or \hii\ emission spectroscopy. While the maximum \hi\ column density 
obtained in the Leiden-Argentine-Bonn (LAB) survey (assuming optically-thin \hii\ 
emission) was $\nhi \approx 2 \times 10^{22}$~\cm\ \citep{kalberla05}, significantly higher $\nhi$ 
values have been inferred from recent modelling of \hii\ emission data of external galaxies 
\citep{braun09,braun12} as well as Lyman-$\alpha$ absorption studies of high-redshift (and 
low-metallicity) gamma ray bursts and quasars \citep{fynbo09,noterdaeme12b}. This raises the question 
of whether the gas is indeed predominantly molecular at $\nhi > 10^{22}$~\cm\ in the Galaxy 
\citep[e.g.][]{schaye01} or whether atomic hydrogen can exist at significantly higher \hi\ 
column densities. In this {\it Letter}, we propose a different approach to determine both the 
\hi\ column density along a sightline and the error on the measurement, based on a combination 
of \hii\ emission and \hii\ absorption spectroscopy. 

\section{The formalism}
\label{sec:formalism}

For the \hii\ line, the two observables are the \hii\ brightness temperature $\Tb$, measured 
from {\it emission} spectroscopy, and the \hii\ optical depth $\tau$, measured from {\it 
absorption} studies towards background radio continuum sources. For a single homogenous 
\hi\ cloud, the observed brightness temperature is given by\footnote{Note that all quantities
in this section are functions of velocity.}
\begin{equation}
\label{eqn:tbts}
\Tb = \Ts \times \left[ 1 - \exp(-\tau) \right] \: \: ,
\end{equation}
while the \hi\ column density $\nhi$, the \hii\ optical depth $\tau$ and the 
spin temperature $\Ts$ are related by the expression 

\begin{equation}
\label{eqn:nhitstau}
\nhi = 1.823 \times 10^{18} \times \int \Ts \: \tau \: {\rm dV} \: ,
\end{equation}
where $\nhi$ is in \cm, the spin and brightness temperatures are in K, and ${\rm dV}$ is in \kms, with 
the integral over the line profile. Note that the above equations make no approximations, 
except that $\Ts >> h \nu_{\rm 21cm}/k_B \approx 0.07$~K \citep[e.g.][]{field58}, which 
should be valid in all astrophysical circumstances. 

For multiple \hi\ clouds along a sightline, equation~(\ref{eqn:nhitstau}) remains unchanged, except that 
$\Ts$ is then the column-density-weighted harmonic mean of the spin temperatures of the different 
clouds along the sightline; we will denote this harmonic mean spin temperature as $\langle \Ts \rangle$. 
However, the expression for $\Tb$ is much more complicated in this situation \citep[e.g.][]{heiles03a}:
\begin{equation}
\label{eqn:tbts-mult}
\Tb = \Sigma_{i=0}^{N-1} T_{s,i} \left[ 1 - \exp(-\tau_i) \right] \times \exp({\left[-\Sigma_{j=0}^{M_i-1} \tau_j \right]}) \:\:,
\end{equation}
for $N$ ``clouds'' with different spin temperatures and optical depths along the sightline,
and $M_i$ clouds between us and the $i$'th cloud.

If the \hii\ absorption is optically thin (i.e. peak optical depth $<< 1$), the above 
expressions can be combined to obtain
\begin{equation}
\label{eqn:nhiot}
\nhiot = 1.823 \times 10^{18} \times \int \Tb {\rm dV} \:\: ,
\end{equation}
and one can estimate the \hi\ column density directly from the \hii\ emission spectrum.
However, in the general case of arbitrary optical depth, it is not possible to determine $\nhi$, 
even on combining the \hii\ absorption and emission spectra. In such a situation, one would, 
in order to determine $\nhi$, need to know both the parameters ($\Ts$, $\tau$) of individual clouds 
as well as the spatial distribution of these clouds along the sightline; the latter is  especially 
difficult to ascertain observationally. 


Our aim is to estimate the \hi\ column density, given measurements of both the brightness 
temperature and the \hii\ optical depth. For this purpose, we define a quantity ${\rm T_{s,eff}}$,
akin to the spin temperature, by the relation
\begin{equation}
\label{eqn:tbtseff}
\Tb = {\rm T_{s,eff}} \times \left[ 1 - \exp(-\tau_{\rm tot}) \right] \: \: ,
\end{equation}
where $\Tb$ is the observed brightness temperature and $\tau_{\rm tot}$ is the {\it total} optical 
depth. We can, without loss of generality, relate the harmonic-mean spin temperature $\langle \Ts \rangle$ 
along a sightline to ${\rm T_{s,eff}}$ by
\begin{equation}
\label{eqn:tstseff}
\langle \Ts \rangle  = f(\Tb,\tau_{\rm tot}) \times {\rm T_{s,eff}}  \:\: ,
\end{equation}
where $f(\Tb,\tau_{\rm tot})$ is some unknown function of $\Tb$ and $\tau_{\rm tot}$ (and which also
depends on the spatial distribution of clouds along the sightline). On replacing in 
equation~(\ref{eqn:nhitstau}) for $\nhi$, we obtain
\begin{equation}
\label{eqn:nhitseff}
\nhi = 1.823 \times 10^{18} \int f(\Tb,\tau_{\rm tot}) \times \frac{\tau_{\rm tot} \times \Tb}
{\left[ 1 - \exp(-\tau_{\rm tot}) \right]} {\rm dV} \:\: .
\end{equation}
Knowledge of the function $f(\Tb,\tau_{\rm tot})$ would allow us to infer the \hi\
column density from measurements of $\Tb$ and $\tau_{\rm tot}$. Of course, this function depends 
on the details of the sightline. However, based on the simulations of the next section, we find
that the median value of the function is $\approx 1$, even in the extreme cases of high $\tau_{\rm tot}$
and high $\nhi$, with a spread of only a factor of a few around the central value. We hence set 
$f \approx 1$ to obtain 
\begin{equation}
\label{eqn:nhits}
\nhits = 1.823 \times 10^{18} \int \frac{\tau_{\rm tot} \times \Tb} {\left[ 1 - \exp(-\tau_{\rm tot}) \right]} {\rm dV} \:\: .
\end{equation}
The above equation to estimate the \hi\ column density was earlier proposed by \citet{dickey82}, 
but has not received much attention in the literature. \citet{dickey82} refer to $\nhits$ as the 
``isothermal'' estimate of the \hi\ column density, since equation~\ref{eqn:nhits} is the same as 
the expression for the \hi\ column density when the \hii\ absorption and emission arise in a single 
cloud with a fixed spin temperature. We will continue to use this terminology, but note that it
can also be regarded as a ``thin-screen'' estimator as it only uses the total optical depth and 
is independent of the spatial disposition of the clouds. In effect, knowledge of the total \hii\ 
optical depth and the total brightness temperature $\Tb$ along the sightline allow one to estimate 
the \hi\ column density. Note that at large $\tau$, equation~(\ref{eqn:nhits}) reduces to 
\begin{equation}
\label{eqn:nhits_high}
\nhits = 1.823 \times 10^{18} \int ({\tau_{\rm tot} \times \Tb}) \: {\rm dV} \:\: .
\end{equation}

\section{Monte Carlo simulations}
\label{sec:sim}

In this section, we use a Monte Carlo procedure to validate the ``isothermal'' estimator 
by measuring the difference between the true \hi\ column density and the \hi\ column densities 
inferred from the isothermal (and optically-thin) methods. The approach taken is the converse of 
the situation in the real world, where we have the measured brightness temperature 
and \hii\ absorption profiles, and would like to infer the \hi\ column density.
Instead, for the purposes of the simulation, we will assume that a sightline contains 
some fiducial ``true'' \hi\ column density $\nhi$ in a narrow velocity channel, that produces
both \hii\ emission and absorption. We further assume that the \hii\ emission and absorption 
in this velocity bin arise from the superposition of contributions from different temperature 
phases along the sightline (each, of course, with an \hi\ column density lower than the total 
\hi\ column).  We distribute the total \hi\ column randomly between the different temperature 
phases, and, for each distribution, compute the observed brightness temperature and the observed 
total optical depth. The inferred $\Tb$ and $\tau_{\rm tot}$ values are then used to determine 
the \hi\ column density from the isothermal and optically-thin methods, and the results compared 
to the known \hi\ column density. This procedure is carried out for a large number of possible
sightlines, covering a wide range of both total \hi\ column densities and distributions of the 
\hi\ column between different temperature phases.

Although we are mainly interested in the high \hi\ column density range, $\nhi \gg 10^{21}$~\cm,
the simulation has been carried out for total $\nhi$ values in the range $10^{20}$~\cm~$\le \nhi 
\le 10^{24}$~\cm. The upper end of the range was chosen so as to probe the extremely high \hi\ 
columns ($\gtrsim 10^{23}$~\cm) whose presence has been suggested by modelling studies of \hi\ 
in local galaxies \citep{braun09,braun12}. The channel width is assumed to be 1~\kms, similar 
to the velocity resolution of the LAB all-sky \hii\ emission survey \citep{kalberla05}. 
This narrow velocity width was chosen to match the narrowest full-width-at-half-maximum of \hii\ 
lines, for purely thermal line broadening of gas at a kinetic temperature of $\approx 20$~K, 
so that there is no loss of information due to under-sampling of the line profiles. Note that 
\hi\ column densities $\gtrsim 10^{23}$~\cm\ are extremely large for such a small velocity 
range; we will discuss the effect of this assumption later. 

We also note that the simulation explicity deals with the \hi\ column density within a 
single {\it observed} 1~\kms\ velocity channel, instead of over some physically motivated 
velocity width. It is well known that \hi\ in the ISM has velocity structure (i.e. correlations 
between neighbouring velocity channels), due to the gas temperature, turbulent motions, 
bulk motions, etc., and the net \hi\ column density along a sightline is inferred from the full 
\hii\ line profile, not merely from individual velocity channels. Our simulation makes no assumptions 
about the velocity structure, because quantities like the gas temperature, turbulent motions and 
velocity gradients are, in general, poorly known, and assumptions about these quantities could bias 
the results. Our approach of using narrow ($\lesssim 1$~\kms) channels does not lead to any loss of 
generality, as the channel width is narrow enough to properly sample the velocity profile. As 
discussed in more detail later, the primary consequence of not using the velocity structure in the 
line profile is that the error bars that we advocate are conservative; the advantage is that no
assumptions are needed about quantities that are difficult to measure.

The next step is to specify the {\it spin} temperatures of the different gas clouds along the 
sightline. In the classic models of the ISM, \hi\ is expected to stably exist in two phases in 
pressure equilibrium with each other \citep[e.g.][]{field69,wolfire03}; these are the 
cold neutral medium (CNM, with $40$~K~$ \lesssim \Tk \lesssim 200$~K) and the warm neutral medium 
(WNM, with $5000$~K~$\lesssim \Tk \lesssim 8000$~K). Recently, there has been evidence that significant 
fractions of \hi\ may be in a thermally unstable phase, with $200$~K~$ \le \Tk \le 5000$~K
\citep[e.g.][]{heiles03a,kanekar03b} and that the CNM temperature range extends to $\approx 20$~K
\citep{heiles03b}. We hence allow for gas in three temperature phases, the CNM, the WNM and 
the thermally unstable neutral medium (UNM), and will also use the temperature range 
$20$~K~$\le \Tk \le 200$~K for the CNM. As will be seen, the precise choice of 
the CNM, WNM and UNM temperature ranges do not significantly affect our results.

To estimate the brightness temperature and the \hii\ optical depth, we require the spin 
temperature, not the kinetic temperature. For the CNM, the \hii\ transition is expected to 
be thermalized by collisions, causing $\Ts \approx \Tk$; we will hence use the spin 
temperature range $20$~K~$\le \Ts \le 200$~K for the CNM. In the case of the WNM (and, possibly, 
the UNM), the low particle number density means that collisions cannot thermalize the line; 
the spin temperature is expected to typically be lower than the kinetic temperature in this phase 
unless there are sufficient Lyman-$\alpha$ radiation in the cloud for resonant scattering of 
these photons to couple the spin and kinetic temperatures \citep[e.g.][]{field58,liszt01}. We have 
hence assumed that $\Ts < \Tk$ in the WNM, and have used the results of \citet{liszt01} to relate 
the spin temperature to the kinetic temperature; this essentially meant using the range 
$2000$~K~$\le \Ts \le 5000$~K for the WNM. The UNM spin temperature range is set to 
$200$~K~$\le \Ts \le 2000$~K, intermediate between the CNM and WNM phases. Note that this 
procedure effectively allows for gas at all spin temperatures $20 \le \Ts \le 5000$~K.


In order to divide the total \hi\ column density into different components along the sightline, 
we first fix the fraction of gas in the three temperature phases. The simulations were carried 
out for the following six cases:

\begin{enumerate}
\item Half the gas is in the CNM, and half in the WNM,

\item Each of the three phases contains one-third of the gas,

\item 50\% CNM, 40\% WNM and 10\% UNM,

\item 50\% CNM, 10\% WNM and 40\% UNM, 

\item 90\% CNM and 10\% WNM, and

\item The fraction in each phase varies randomly from run to run.

\end{enumerate}

Having fixed the total \hi\ column density and the gas fraction in each phase,
we know the \hi\ column density in each phase. We allow for the possibility that there
could be multiple ``clouds'' from a given phase that contribute to the
emission/absorption in the velocity channel by allowing gas in each phase
to be further sub-divided into multiple components. We then randomly assign a spin
temperature to the first component of each phase, subject to the constraint that it lies
within the $\Ts$ range of the phase. The \hi\ column density of the component is also
randomly chosen from the range $1\times 10^{19}$~\cm\ to $5 \times 10^{22}$~\cm. We then
add new components of the same phase until the sum of their \hi\ column densities is equal
to or greater than the total \hi\ column density set for the phase. In the latter case, the
\hi\ column density of the last component is reduced to a value which makes the sum of
\hi\ column densities match the total \hi\ column density. In this process, if the 
$\nhi$ of the last component falls below the minimum allowed value ($1\times 10^{19}$~\cm), 
all the components are discarded and the entire process is repeated. Similarly, if the 
brightness temperature along the line of sight exceeds 500~K, then all components are 
discarded and the process is repeated. Note that the highest brightness temperature 
observed in the Galaxy is $\approx 200$~K, significantly lower than the above threshold.

The above procedure is carried out for all three phases. Once all the components in all 
three phases have been chosen, the order of the components along the sightline is randomized. 
The known \hi\ column density and spin temperature of each component are then used to 
compute the two observables for the sightline, the total optical depth $\tau_{\rm tot}$ and 
the brightness temperature $\Tb$; these are then used in equations (\ref{eqn:nhiot}) 
and (\ref{eqn:nhits}) to estimate the \hi\ column density from the optically-thin ($\nhiot$) 
and isothermal ($\nhits$) methods. Of course, the true \hi\ column density along the sightline, 
$\nhi$, is already known. For each bin in total ``true'' \hi\ column density (bin width~$= 0.235$ 
in $\log[\nhi]$) and total \hii\ optical depth (bin width~$=0.2$), we carried out 2001 runs for each 
of the six distributions between the different temperature phases, computing the statistics of 
the ratios $(\nhi/\nhits)$ and $(\nhi/\nhiot)$ in every case.

Finally, we also separately ran the simulations with the temperature ranges $20$~$\le \Ts \le 200$~K 
for the CNM, $200$~K~$\le \Ts \le 5000$~K for the UNM, and $5000$~K~$\le \Ts \le 8000$~K for the 
WNM, i.e. assuming that the \hii\ transition is thermalized (with $\Ts \approx \Tk$) in all three
phases. This too was done for the above six distributions of gas between the different phases. No 
significant difference was found between the results of the simulations using $\Ts \approx \Tk$ 
and those using the \citet{liszt01} relation between $\Ts$ and $\Tk$; we will hence restrict 
the discussion to the latter in the following sections.

\begin{figure*}
\begin{center}
\includegraphics[scale=0.3,angle=270.0]{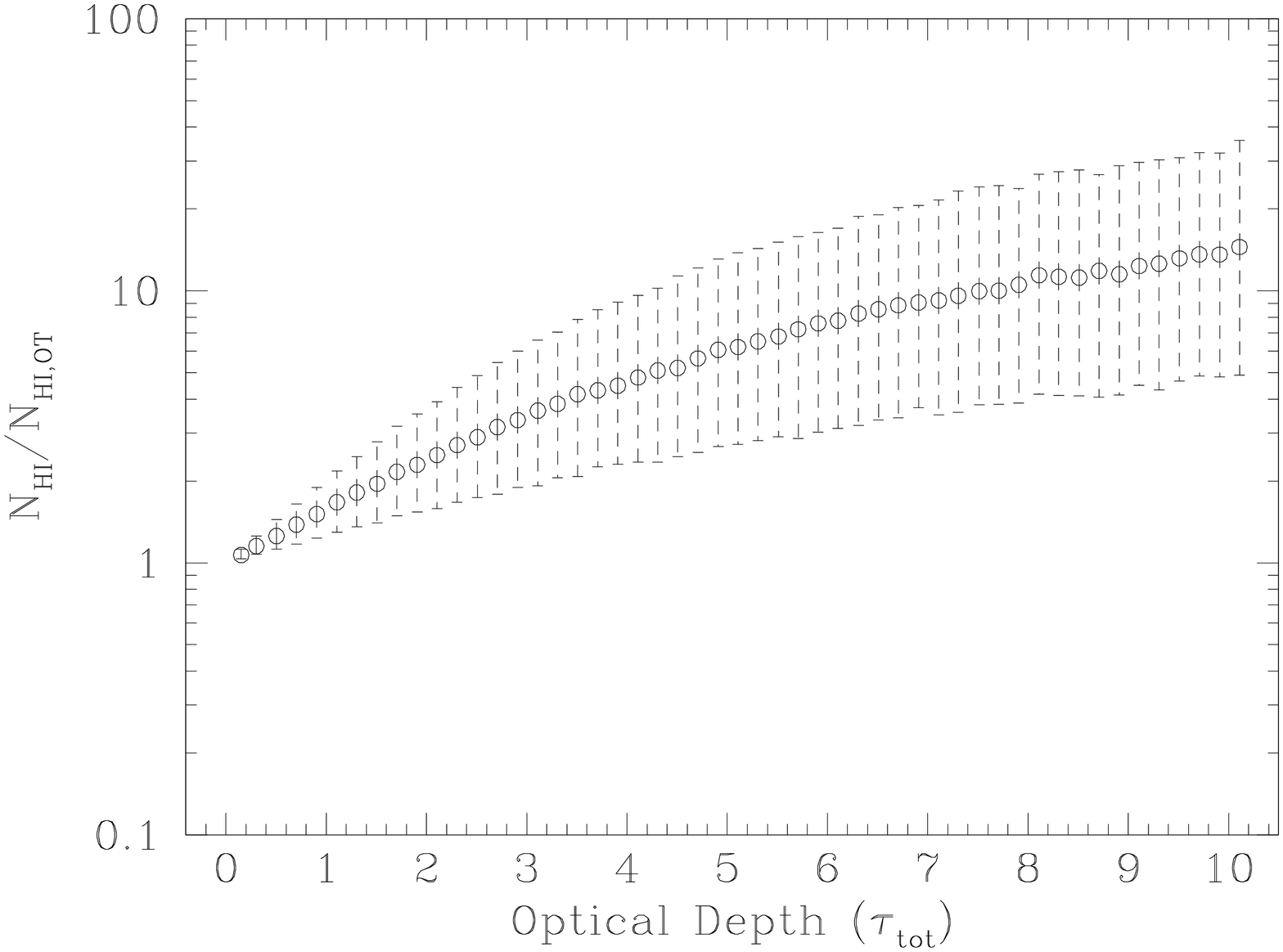}
\includegraphics[scale=0.3,angle=270.0]{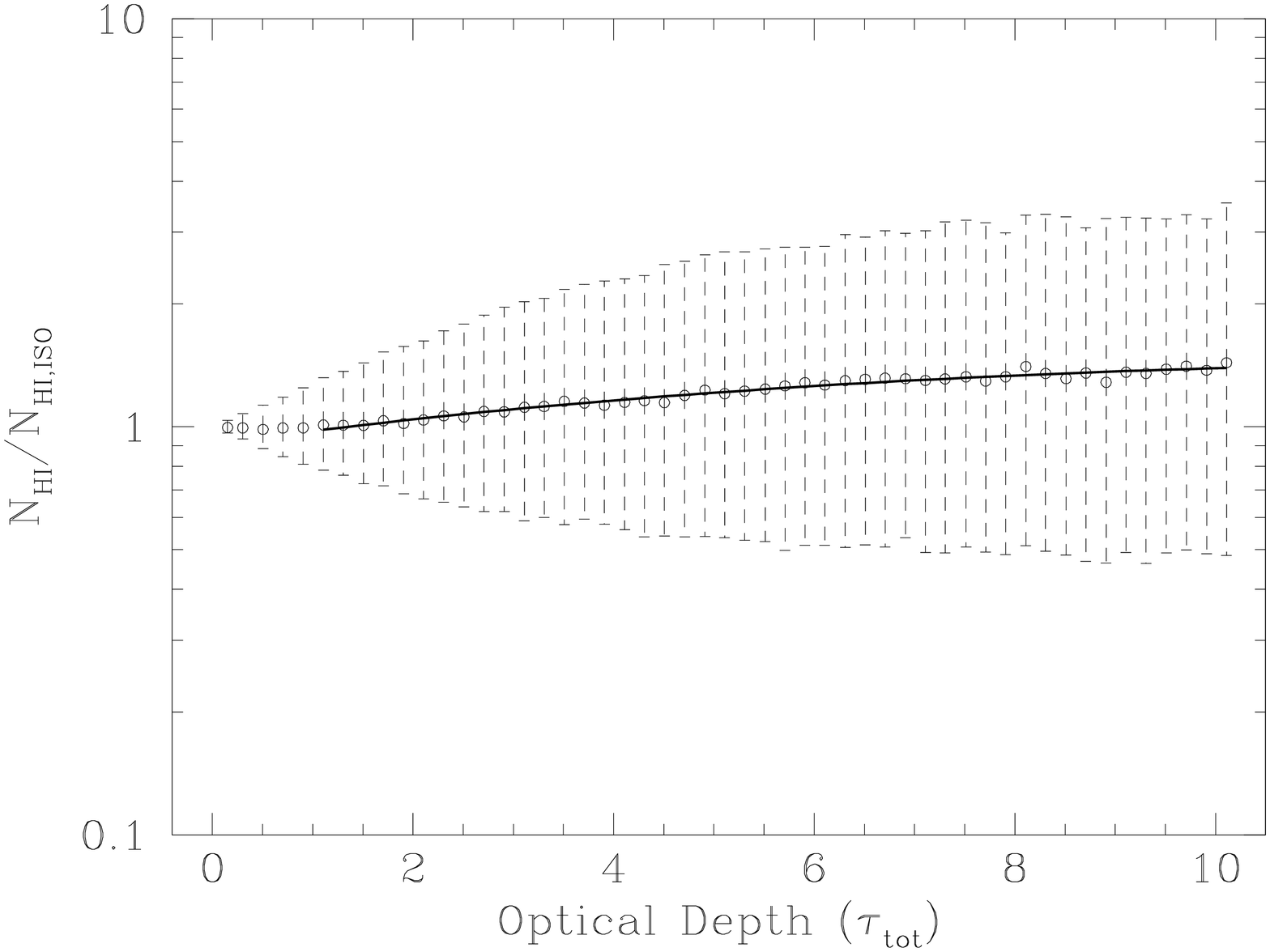}
\caption{The ratio of the true \hi\ column density $\nhi$ to the \hi\ column 
density obtained from [A]~the optically-thin estimate $\nhiot$ (left panel) 
and [B]~the isothermal estimate $\nhits$ (right panel), plotted against total optical depth 
$\tau_{\rm tot}$ (with a bin width of 0.2). In both panels, the filled circles mark 
the median value of the ratio, while the error bars encompass the 68.2\% confidence intervals. 
For each bin, the median and the confidence interval were computed over a total of 2001 runs.
The simulation is for a random distribution of gas between the three temperature phases, 
using the \citet{liszt01} relation between spin and kinetic temperatures. The solid 
line in the right panel shows the best 2nd-order polynomial fit to the relation between
$\left[\nhi/\nhits\right]$ and $\tau_{\rm tot}$, valid for $\tau_{\rm tot} > 1$: 
$\left[\nhi/\nhits\right] = 0.904 + 0.074 \tau_{\rm tot} - 0.0026 \tau^2_{\rm tot}$. }
\label{fig:ratio_tau} 
\end{center}
\end{figure*}

\begin{figure*}
\begin{center}
\includegraphics[scale=0.3,angle=270.0]{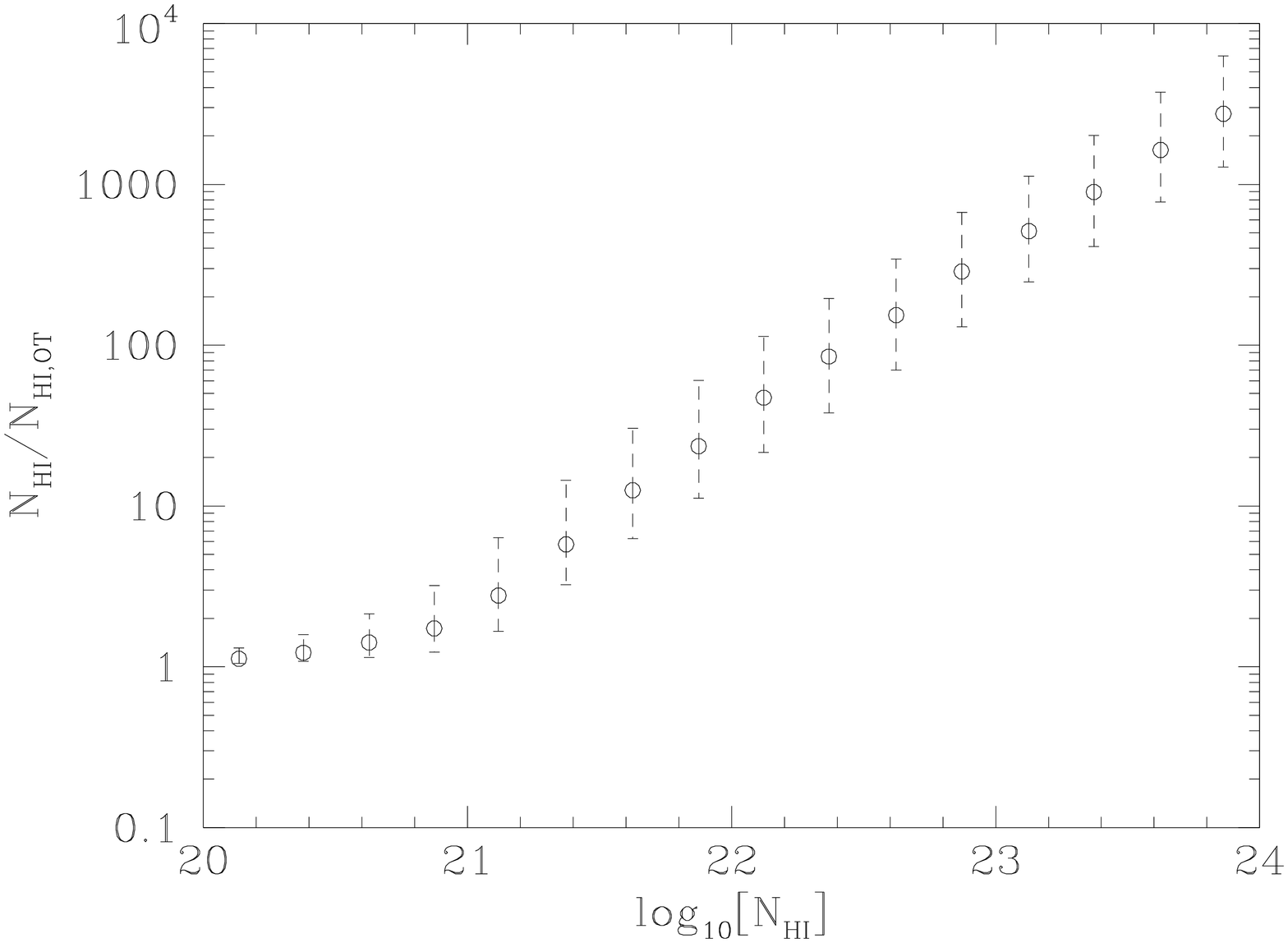}
\includegraphics[scale=0.3,angle=270.0]{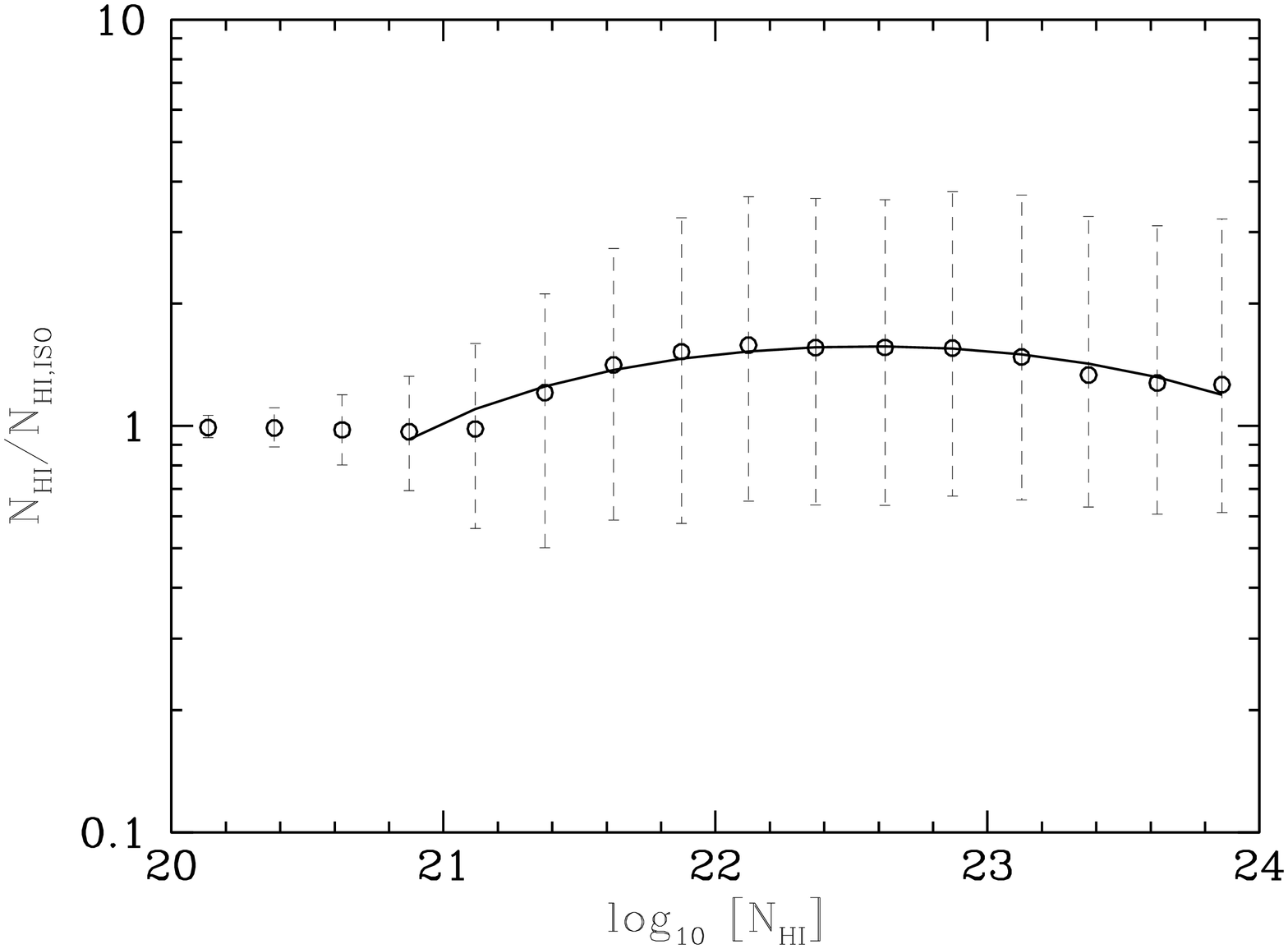}
\caption{The ratio of the true \hi\ column density $\nhi$ to [A]~the optically-thin 
estimate $\nhiot$ (left panel) and [B]~the isothermal estimate $\nhits$ (right panel), 
plotted against $\log[\nhi]$ (with a bin width of $0.235$ in $\log[\nhi])$. 
In both panels, the filled circles mark the median value, while the error bars give
the 68.2\% confidence intervals. For each bin, the median and the confidence interval were 
computed from 2001 runs. The simulation is for a random distribution of gas between the 
three temperature phases, using the \citet{liszt01} relation between spin and kinetic temperatures. 
The solid line in the right panel shows the best 2nd-order polynomial fit to the relation between
$\left[\nhi/\nhits\right]$ and $\log[\nhi]$, valid for $\nhi \geq 10^{21}$~\cm: 
$\left[\nhi/\nhits\right] = -112.88 + 10.144 \times \log[\nhi/{\rm cm}^{-2}] - 0.2248 \times \log[\nhi/{\rm cm}^{-2}]^2$. }
\label{fig:ratio_nhi} 
\end{center}
\end{figure*}

\section{Results and Discussion}
\label{sec:res}

In practice, qualitatively similar results were obtained for the different distributions 
of gas between the different temperature phases. We hence show results only for the 
last case, where the fraction of gas in the CNM, WNM and UNM phases is allowed to vary 
randomly from run to run. This is likely to provide the most robust estimate 
of the error on the results, unbiased by a specific choice of a gas distribution.

The critical quantity in determining the accuracy of the optically-thin and isothermal 
estimates of the \hi\ column density is the total optical depth along the sightline 
$\tau_{\rm tot}$. The accuracy of either estimate is worse at high optical depths than for 
$\tau \lesssim 1$, because it is easier to hide \hi\ behind foreground gas that is optically 
thick.  Figs.~\ref{fig:ratio_tau}[A] and [B] show, respectively, the ratio of the true \hi\ column 
density to the optically-thin and isothermal estimates of $\nhi$, for the case where the 
fractions of gas in the three temperature phases are allowed to vary randomly from run to 
run. As expected, the optically-thin estimate always yields a lower limit to the true \hi\ 
column density, i.e. the ratio is always $> 1$. Even for $\tau_{\rm tot} \approx 1$, the median 
\hi\ column density derived from the optically-thin estimate underestimates the true \hi\ 
column density by a factor of $\approx 1.6$; the under-estimate is by more than an order of 
magnitude for optical depths of $\approx 10$. The spread on the optically-thin estimate 
is also very large, with the 68.2\% confidence level on the ratio $\nhi/\nhiot$ reaching 
$\approx 40$ for $\tau_{\rm tot} \approx 10$.  

Conversely, the median isothermal estimate of the \hi\ column density $\nhits$ tracks the true 
\hi\ column density to better than 10\% even for optical depths of $\approx 5$. Even for 
$\tau_{\rm tot} \approx 10$, the median value of $\nhi/\nhits$ is within a factor of 1.5 of unity. 
The 68.2\% confidence intervals on the ratio extend from $\approx 0.4$ to $\approx 3$, even at 
high optical depths. Clearly, even for $\tau_{\rm tot} \approx 10$, the isothermal estimate 
of the \hi\ column density appears accurate to within a factor of $\approx 3$. 

Figs.~\ref{fig:ratio_nhi}[A] and [B] show the same ratios plotted in Fig.~\ref{fig:ratio_tau}, 
but this time as a function of the true \hi\ column density. Note that the largest total optical 
depth on sightlines included in this figure is $\approx 1000$. The left panel [A] shows that, 
while the optically-thin estimate provides a lower limit to the true $\nhi$, this is not 
particularly useful at large values of $\nhi$ ($\gtrsim 10^{22}$~\cm), for which the true 
$\nhi$ is more than an order of magnitude larger than $\nhiot$. On the other hand, the median 
value of the thin screen estimate $\nhits$ tracks the true $\nhi$ to within a factor of 2 
over the entire \hi\ column density range ($\nhi \lesssim 5 \times 10^{23}$~\cm). Further,
even the 68.2\% confidence level intervals of the ratio lie within a factor of a few from
unity. Clearly, the scatter in the estimated $\nhits$ is quite modest even for very high \hi\ 
column densities.

The simulation results thus indicate that the isothermal estimate of equation~(\ref{eqn:nhits})
provides a fairly good measure of the true \hi\ column density, certainly good to within a 
factor of a few within 68.2\% confidence intervals, even for high optical depths,
$\tau_{\rm tot} \approx 1000$, and high \hi\ column densities, $\nhi \approx 5 \times 10^{23}$~\cm.
(Note that optical depths of $\approx 1000$ are significantly higher than would be expected along 
Galactic sightlines.) This estimator should thus allow one to directly probe the existence of 
very high \hi\ column densities ($\nhi \gtrsim 10^{23}$~\cm) in the Galaxy, via accurate measurements 
of the \hii\ brightness temperature and optical depth from, respectively, high velocity resolution 
\hii\ emission and absorption spectroscopy. The best fit 2nd-order polynomials to the ratio of the
true \hi\ column density to the isothermal estimate of the \hi\ column density are given in the 
captions to Figs.~\ref{fig:ratio_tau} and \ref{fig:ratio_nhi}. Note that these are valid for 
$\tau_{\rm tot} \gtrsim 1$ and $\nhi \gtrsim 10^{21}$~\cm, respectively. Of course, at lower optical depths,
$\tau_{\rm tot} < 1$, the optically-thin estimate provides an acceptable measure of the true 
\hi\ column density, within a factor of $\approx 1.5$ of the true value.

It should be emphasized that the isothermal approach assumes implicitly that the \hii\ brightness 
temperature and optical depth are measured {\it along the same sightline}. Of course, this
is not the case in reality, with the optical depth along a sightline usually measured towards 
compact background radio sources and the brightness temperature inferred 
by interpolating between measurements at neighbouring positions \citep[e.g.][]{heiles03a}.
Small-scale structure in the \hi\ could imply incorrect brightness temperature estimates 
along the sightline and hence larger errors in the estimate of the \hi\ column density. 
We do not anticipate that this will be a severe problem for most sightlines, especially given the 
possibility of obtaining \hii\ emission spectra with telescopes of very different angular 
resolutions (e.g. from the LAB survey, the Green Bank Telescope, the Arecibo telescope, etc).

The approach also assumes that it is possible to accurately measure very high \hii\ optical depths, 
$\tau_{\rm tot} >> 10$. In reality, {\it direct} measurements of such opacities would require 
spectral dynamic ranges $\gg 10^{4}$ per 1~\kms\ channel, that will not 
be achieved even with future facilities like the Square Kilometer Array. However, it should be possible 
(except for cases with severe line blending) to determine the opacity at line peak via a Gaussian-fitting 
procedure to spectra of a high velocity resolution, as the wings of the features (which have lower 
optical depths) would also have been measured at very high signal-to-noise ratios. Note that the 
highest \hii\ opacities would arise for cold gas with high \hi\ column density, which is likely to 
be thermalized and, hence, to have a Gaussian line profile. We have verified from simulations that 
opacities of $\gtrsim 50$ can be easily measured via such a Gaussian-fitting procedure, from spectra 
with velocity resolutions of $\approx 0.5$~\kms\ and optical depth RMS noise values of $\approx 0.001$ 
per 0.5~\kms\ channel \citep[which have already been achieved with today's interferometers; e.g.][]{braun05}.

The present simulations have been carried out so as to match observations with high velocity resolution, 
$\lesssim 1$~\kms. This is to ensure that there is no loss of information due to smoothing of narrow 
absorption from cold \hi\ clouds, even at $\Tk \approx 20$~K. The quoted errors above on the inferred 
\hi\ column density are per 1~\kms\ spectral channel. However, as noted earlier, the velocity width of 
absorption/emission from individual gas ``clouds'' with $\Tk \gg 20$~K will always be significantly 
larger than 1~\kms. As a result, the error introduced when estimating the total \hi\ column density along 
a sightline by integrating the per channel isothermal estimate across the line profile will be correlated 
between groups of neighbouring 1~\kms\ channels. A conservative estimate of the net systematic error on 
the total \hi\ column density along the sightline can be obtained by assuming that the errors are perfectly 
correlated over the line profile, implying that the error on the total \hi\ column density is the same as 
that one on the \hi\ column density per 1~\kms\ velocity channel, as determined from the present simulation.

We also note that the assumption of \hi\ column densities of $\approx 10^{23} - 10^{24}$~\cm\ per 
1~\kms\ velocity channel is quite extreme. Extremely high \hi\ column densities, $\gtrsim 10^{23}$~\cm,
are more likely to be distributed over a significantly larger velocity range, with far smaller columns
arising per 1~\kms\ channel. Allowing for such large column densities per 1~\kms\ channel results 
in both shifting the median value of the ratio $\nhi/\nhits$ slightly away from unity as well as 
significantly increasing the error bars in the isothermal estimate. For example, if we limit to 
$\nhi = 10^{22}$~\cm\ per 1~\kms\ velocity channel, the median $\nhi/\nhits$ is $\le 1.2$ for $\tau_{\rm tot}
< 10$, while the 68.2\% confidence level intervals on the ratio extend from $\approx 0.5$ to $\approx 2$.
Our choice of $10^{24}$~\cm\ as the limiting \hi\ column density per 1~\kms\ channel is thus a very
conservative one.

Finally, other estimators of the \hi\ column density have been used in the literature 
\citep[e.g.][]{lockman95,wakker11}, albeit without a detailed characterization of their 
accuracy in the high $\nhi$ regime. For example, \citet{wakker11} use an estimator that appears 
similar to the isothermal estimate, with 
\begin{equation}
\label{eqn:wakker}
\nhi = 1.823 \times 10^{18} \times \int \Ts \log \left[ \frac{\Ts}{\Ts - \Tb} \right] {\rm dV} \: ,
\end{equation}
and assume $\Ts = 135$~K for the half of the spectrum with the highest $\Tb$ values and $\Ts = 5000$~K 
for the half of the spectrum with the lowest $\Tb$ values. While this has the advantage that only 
the \hii\ emission spectra are needed to determine the \hii\ column density (as is also true for 
the optically-thin estimate), it is not very useful at high optical depths, where $\Ts \approx \Tb$. 
It is in the high opacity regime that the isothermal estimate is clearly superior to the other methods.

\section{Summary}
\label{sec:summary}

We study in detail the ``isothermal'' estimator of the \hi\ column density, earlier proposed by \citet{dickey82},
based on measurements of the total \hii\ brightness temperature $\Tb$ and the total \hii\ optical depth 
$\tau_{\rm tot}$ from high velocity resolution spectroscopy. The ``isothermal'' estimate is given by 
$\nhits = 1.823 \times 10^{18} \int \left[ \tau_{\rm tot} \times \Tb \right] /  \left[ 1 - e^{-\tau_{\rm tot}} 
\right] {\rm dV}$. We have carried out a Monte Carlo simulation of realistic sightlines, including gas in 
different phases and random locations along the path, to determine the accuracy of the isothermal estimate of 
the \hi\ column density. We find that this approach yields accurate estimates of the \hi\ column density, 
and that the results do not strongly depend (statistically) on the spatial distribution of gas along the sightline,
or its distribution in different temperature phases. In general, the median isothermal estimate of the \hi\ 
column density is within a factor of 2 of the true \hi\ column density, while the 68.2\% confidence intervals 
of the ratio of $\nhi/\nhits$ are within a factor of $\approx 3$ of unity, out to extremely high \hi\ column 
densities, $\le 5 \times 10^{23}$~\cm\ per 1~\kms\ channel, and high total optical depths, $\le 1000$. The 
68.2\% confidence intervals are conservative, as they allow for the possibility of extremely high \hi\ 
column densities, $> 10^{23}$~\cm, in a narrow velocity range (1~\kms), which is unlikely to arise in 
reality. We conclude that the isothermal estimator allows one to accurately measure the \hi\ column 
density (to within a factor of a few even at the highest \hi\ column densities) and to thus directly 
probe the existence of extremely high \hi\ column density gas in the Galaxy.

\section*{Acknowledgements}

NK acknowledges support from the Department of Science and Technology through a 
Ramanujan Fellowship. NR acknowledges the Jansky Fellowship Program of NRAO/NSF/AUI and
support from the Alexander von Humboldt Foundation. We thank Robert Braun and an anonymous 
referee for comments on an earlier version of this paper, and Harvey Liszt for providing
us with a table giving the relation between spin and kinetic temperatures in his model.

\label{lastpage}

\bibliographystyle{mn2e}
\bibliography{ms}

\end{document}